\documentstyle[pra,twocolumn,aps]{revtex}

\begin{document}

\title{Einstein, Podolsky, Rosen, and Shannon\footnote{Dedicated to the
memory of James T. Cushing.}}

\author{Asher Peres}
\address{Department of Physics, Technion---Israel Institute of
Technology, 32000 Haifa, Israel}

\maketitle
\begin{abstract}
The Einstein-Podolsky-Rosen paradox (1935) is reexamined in the light
of Shannon's information theory (1948). The EPR argument did not take
into account that the observers' information was localized, like any
other physical object.

\end{abstract}

\vspace*{3cm}
I had the privilege of meeting Jim Cushing in 1986, during a conference
that Daniel Greenberger had organized in the World Trade Center in New
York City \cite{nyas}. Some time later Cushing sent me a copy of his
book \cite{book}, where our encounter is related in a footnote:

\begin{quote}{``I mentioned to Peres that his position appeared to be an
instrumentalist one. He replied with no apparent discomfort that others
had told him that before. For a physicist's statement on an
instrumentalist interpretation of quantum mechanics, see Peres (1988)
\cite{cat}.''}\end{quote}

\noindent Jim had asked me that question with the same tone as if he were
asking whether I was a cannibal. For a more recent discussion on this
subject, see \cite{opinion}.

On the other hand, I never had the privilege of meeting Albert Einstein.
He died when I was an undergraduate. I had always been fascinated by
Einstein, as any normal Jewish boy would be, and later in my life I
even got the impression that I came to know him personally. This is
because my PhD thesis advisor was Nathan Rosen, who had been a close
collaborator of Einstein. Together they built the Einstein-Rosen bridge
in General Relativity \cite{bridge}, and together with Podolsky they
formulated the famous EPR paradox \cite{epr}. Rosen told me many
anecdotes about Einstein and his reactions to various events. (Rosen's
wife Hanna, who was an accomplished pianist, gave piano accompaniment
to Einstein who played the violin.)

My first encounter with the EPR paper occurred when I was a graduate
student, circa 1958. The subject of my research was gravitational
radiation. At that time, I was rather ignorant of quantum theory, having
graduated in mechanical and nuclear engineering, not in physics. One
day, I came into Rosen's office, and I found him sorting out his
papers. On the ground, there were cartons full of old reprints, with
the characteristic green covers of The Physical Review. One of them
read: ``Can Quantum-Mechanical Description of Physical Reality Be
Considered Complete?'' \,The authors were A.~Einstein, B.~Podolsky,
and N.~Rosen.  I thought that this would be a nice item to have in my
collection of reprints, and politely asked:

``Professor Rosen, may I take one of these reprints?''

\noindent He looked concerned.

``Ha, how many are left?''

\noindent We counted them, which was not difficult:
there were two. Then he said, with some hesitation,

``Well, if there are two, you may have one.''

\noindent This is how I acquired the last available reprint of the
famous article of Einstein, Podolsky, and Rosen.

Let us have a look at that wonderful paper. You will immediately notice
that Eqs.~(7) and~(8) involve {\it entangled\/} wave-functions, and
indeed the whole issue is about the physical consequences of such an
entanglement. Entangled wave-functions were not new at that time: you
can find one, for example, in Eq.~(10) of Rosen's 1931 seminal paper on
the ground state of the hydrogen molecule~\cite{H2}, which is probably
more famous among chemists than the EPR paper is among physicists.

Some time after that work, Rosen became a post-doc of Einstein at the
Institute of Advanced Studies in Princeton. One day, at the traditional
3~o'clock tea, Rosen mentioned to Einstein a fundamental issue of
interpretation related to entangled wave-functions. Einstein immediately
saw the implications for his long standing disagreement with Bohr. As
they discussed the problem, Boris Podolsky joined the conversation,
and later proposed to write an article. Einstein acquiesced. When he
later saw the text, he disliked the formal approach, but agreed to
its publication. Then, as soon as the EPR article appeared, Podolsky
relased its contents to the New York Times (4 May 1935, page 11) in
a way implying that the authors had found that quantum mechanics was
faulty. This infuriated Einstein, who after that no longer spoke
with Podolsky.

The EPR ``paradox'' drew immediate attention. Niels Bohr \cite{bohr}
found the reasoning faulty, because it contradicted his complementarity
principle. Bell, in his first article on hidden variables and
contextuality~\cite{bell66}, wrote ``the Einstein-Podolsky-Rosen
paradox is resolved in the way which Einstein would have liked least."
Actually, the example given by Bell in the proof of his celebrated
theorem~\cite{bell64} is based on a much simpler entangled system:
two spin-$1\over2$ particles in a singlet state~\cite{bohm}.  In 1991,
David Mermin came to Technion to give the annual Wunsch Lecture. He
said that the EPR paper was wrong. After the talk, there was as usual
a discussion period. Nathan Rosen politely commented: the paper is
not wrong, it makes some assumptions, and then draws the logical
conclusions; the assumptions were wrong.

The EPR article was not wrong, but it had been written too early.
Only some years later, in 1948, Claude Shannon published his theory
of information~\cite{shannon} (and it took many more years before
the latter was included in the physicist's toolbox). Shannon was
employed by the Bell Telephone Company and his problem was to make
communication more efficient. Shannon showed that information could be
given a quantitative measure, that he called {\it entropy\/}. It was
later proved that Shannon's entropy is fully equivalent to ordinary
thermodynamical entropy~\cite{bennett}. Information can be converted
to heat and can perform work. Information is not just an abstract
notion~\cite{landauer}. It requires a physical carrier, and the latter
is (approximately) {\it localized\/}. After all, it was the business of
the Bell Telephone Company to transport information from one telephone
to another telephone, in a different location.

In the EPR article, the authors complain that ``it is possible to assign
two different wave functions to \ldots\ the second system,'' and then,
in the penultimate paragraph, they use the word {\it simultaneous\/} no
less than four times, a surprising expression for people who knew very
well that this term was undefined in the theory of relativity. Let
us examine this issue with Bohm's singlet model. One observer,
conventionally called Alice, measures the $z$-component of the spin of
her particle and find $+\hbar/2$. Then she {\it immediately\/} knows
that if another distant observer, Bob, measures (or has measured,
or will measure) the $z$-component of the spin of his particle, the
result is certainly $-\hbar/2$. One can then ask: when does Bob's
particle acquire the state with $s_z=-\hbar/2$?

This question has two answers. The first answer is that the question is
meaningless --- this is undoubtedly true. The second answer is that,
although the question is meaningless, it has a definite answer: Bob's
particle acquires this state {\it instantaneously\/}. This then raises a
new question: in which Lorentz frame is it instantaneous? Here, there is
also a definite answer: it is instantaneous in the Lorentz frame that we
arbitrarily choose to perform our calculations~\cite{interv2}. Lorentz
frames are not material objects: they exist only in our imagination.

When Alice measures her spin, the information she gets is localized at
her position, and will remain so until she decides to broadcast it.
Absolutely {\it nothing\/} happens at Bob's location. From Bob's point
of view, all spin directions are equally probable, as can be verified
experimentally by repeating the experiment many times with a large
number of singlets without taking in consideration Alice's results.
Thus, after each one of her measurements, Alice assigns a definite
pure state to Bob's particle, while from Bob's point of view the state
is completely random ($\rho$ is proportional to the unit matrix). It
is only if and when Alice informs Bob of the result she got (by mail,
telephone, radio, or by means of any other material carrier, which is
naturally restricted to the speed of light) that Bob realizes that his
particle has a definite pure state. Until then, the two observers can
legitimately ascribe different quantum states to the same system. For
Bob, the state of his particle suddenly changes, not because anything
happens to that particle, but because Bob receives information about
a distant event. Quantum states are not physical objects: they exist
only in our imagination.

In summary, the question raised by EPR ``Can quantum-mechanical
description of physical reality be considered complete?'' has a positive
answer. However, reality may be different for different observers.

\bigskip This work was supported by the Gerard Swope Fund.

\end{document}